\documentclass[a4,reqno,10pt]{amsart}

 \usepackage{amsmath,amsthm,amssymb,xcolor,graphicx,soul}

\usepackage{amscd,verbatim,a4wide}
\usepackage[all]{xy}

\usepackage{hyperref}

\def\G{\mathcal G}
\def\jp{\frac{1}{2}}
\def\be{\begin{equation}}
\def\ee{\end{equation}}

\def\be{\begin{equation}}
\def\ee{\end{equation}}
\def\D{\mathcal D}
\def\G{\mathcal G}
\def\H{\mathcal {H}}
\def\E{\mathcal E}
\def\bc{\mathbb C}
\def\br{\mathbb R}
\def\R{\mathcal R}
\def\J{\mathcal J}

\def\jp{\frac{1}{2}}
\def\ri{{\mathrm i}}

\newcommand {\supplus}{\mathop{{\supset}\llap{\raise
0.5pt\hbox{\normalfont\small+}\hskip 0.5pt}}}

\newcommand {\subplus}{\mathop{{\subset}\llap{\raise
0.5pt\hbox{\normalfont\small+}\hskip 0.5pt}}}

\theoremstyle{plain}

\theoremstyle{definition}

\theoremstyle{remark}

\numberwithin{equation}{section}
\numberwithin{theorem}{section}
\numberwithin{figure}{section}
\numberwithin{table}{section}

\setstcolor{red}

\usepackage{tikz}
\usetikzlibrary{arrows}
\usetikzlibrary{positioning}
\usepackage{amsmath}
\usepackage{amsfonts}
\usepackage{hyperref}
\usepackage{amsthm}

\begin{document}

\title[Yang-Baxter  $\sigma$-model with WZNW term as ${ \mathcal E}$-model]{Yang-Baxter $\sigma$-model with WZNW term as ${ \mathcal E}$-model}

\author[C Klim\v c\'\i k]{Ctirad Klim\v c\'\i k}
\address[C Klim\v c\'\i k]{
Institut de Math\'ematiques de Luminy,
Aix Marseille Universit\'e, CNRS, Centrale Marseille I2M, UMR 7373,
13453 Marseille, France}
\email{ctirad.klimcik@univ-amu.fr}

\begin{abstract}
It turns out that many integrable $\sigma$-models on group manifolds belong to the class of the so-called $\E$-models which are relevant
in the context of the Poisson-Lie T-duality. We show that this is the case also for the Yang-Baxter $\sigma$-model with WZNW term introduced 
by Delduc, Magro and Vicedo in \cite{DMV15}.
\end{abstract}

\maketitle


\section{Introduction}

Integrability and T-dualisability  are both rare properties  of non-linear $\sigma$-models  and it is not known what necessary conditions must be imposed on a target space metric 
$G$  and  on closed 3-form field $H$  in such a way that the corresponding $\sigma$-model enjoys either one or another.   
It turns out, however, that  those two properties share something in common, in the sense
that the majority of integrable sigma models \cite{int,K02,K09,K14}  recently constructed fall in the class of the so-called $\E$-models \cite{K15}, which are the building-blocks for the construction of T-dualisable models \cite{KS95,KS96}.
It is not yet clear   why it is so, but we find that trying to put all integrable $\sigma$-models  under the common roof  of  the $\E$-models is a nice guiding  principle which may eventually lead to a deeper understanding of this phenomenon.
In particular, the purpose of the present note is to give the $\E$-model interpretation  of one of a few integrable $\sigma$-models for which  this was not yet done, namely  the  Yang-Baxter  $\sigma$-model with WZNW term. This 
model  was introduced in \cite{DMV15} (see also \cite{KOY11} for previous work) and it  is a two parameter\footnote{Since the constant $K$ is just the overall normalisation constant and the relation \eqref{-2} is supposed to hold, the action \eqref{-1} is indeed a two-parameter deformation of the principal chiral model; the latter is recovered for the particular values $\eta=A=k=0$.}   deformation of the principal chiral model
defined by the action 
\be S[g]=-\frac{K}{4}\left(\int d\tau \oint  \left(g^{-1}\partial_-g,(1+\eta^2+AR+\eta^2R^2)g^{-1}\partial_+g\right)  +k\int d^{-1}\oint ( g^{-1}dg,[g^{-1}\partial_\sigma g,g^{-1}dg]) \right).\label{-1}\ee
Here $(.,.)$ is the Killing-Cartan form on  the Lie algebra $\G$ of a simple compact Lie group $G$,   $R:\G\to\G$  is the so-called Yang-Baxter operator (see Section 3 for  its definition), $\tau$ and $\sigma$ stand respectively for the world-sheet  time and space variables (the time
and space integrals are notationally separated as $\int$ and $\oint$ expressing\footnote{Note also our perhaps non-standard way of writing of the Wess-Zumino term which is however very useful for "travelling" between the first order and the second order description of the $\sigma$-model dynamics. Mathematically speaking, the object $\oint ( g^{-1}dg,[g^{-1}dg,g^{-1}\partial_\sigma g])$ should be viewed as a two-form on the infinite dimensional loop group manifold $LG$.  } the angular character of the   variable $\sigma$), $g\equiv g(\sigma,\tau)$ is a dynamical field configuration with values in $G$, we set  also
$\partial_\pm:=\partial_\tau\pm\partial_\sigma$ and, finally, the parameters of the model are constrained by 
the  relation
\be A=\eta\sqrt{1-\frac{k^2}{1+\eta^2}}.\label{-2}\ee
While Delduc, Magro and Vicedo showed in \cite{DMV15} that  the model  
 \eqref{-1}  is integrable only if the relation \eqref{-2} holds, we show in the present paper  that  \eqref{-1}   has the structure of   the $\E$-model  also only in the case when the constraint \eqref{-2}  is  imposed.
 This fact illustrates again the mysterious relation between the $\E$-model formalism and the integrability which motivates our work.

 The plan of this note is as follows : We first review  how the concept of the $\E$-model is built up on that of the Drinfeld double, then we introduce a particular  Drinfeld double  and show that the $\E$-model which corresponds to it
 is just the Delduc, Magro and Vicedo model \eqref{-1}. We also express the Lax  pair of the integrable model \eqref{-1}  in the formalism of the $\E$-model therefore our paper can be viewed also as an alternative demonstration of the integrability 
 of  the model \eqref{-1}.

\section{$\E$-models : generalities}
 
Consider thus  a  real Lie algebre $\D$ equipped with a non-degenerate symmetric ad-invariant bilinear form $(.,.)_\D$ and consider an infinte-dimensional manifold $P$ the points of which are smooth maps from a circle $S^1$ into $\D$.  By  parametrizing the circle by the angle $\sigma\in[0,2\pi]$ we can  associate to each $\sigma$ a $\D$-valued function  $j(\sigma)$ on $P$   which to every map $S^1\to\D$ attributes its value in  $\sigma$. By picking further a basis $T^A$ of $\D$ we can construct also numerical functions $j^A(\sigma)$ on $P$ by the prescription
\be j^A(\sigma):= (j(\sigma),T^A)_\D.\ee
We interpret the numerical functions $j^A(\sigma)$ on $P$ as coordinates on $P$ and we define a Poisson structure $\{.,.\}_P$ on $P$ by the following Poisson brackets of the coordinates :
\be \{j^A(\sigma_1),j^B(\sigma_2)\}=F^{AB}_{\phantom{AB}C}j^C(\sigma_1)\delta(\sigma_1-\sigma_2)+D^{AB}\delta'(\sigma_1-\sigma_2).\label{1}\ee
Here the  tensors $F^{AB}_{\phantom{AB}C}$ and $D^{AB}$ are defined by means of the structure of the Lie algebra $\D$
\be D^{AB}:=(T^A,T^B)_D, \quad [T^A,T^B]=F^{AB}_{\phantom{AB}C}T^C.\label{2}\ee
In fact, the non-degeneracy of the bilinear form $(.,.)_\D$ guarantees that  the Poisson structure $\{.,.\}_P$ on $P$ is   symplectic\footnote{The Poisson structure given by Eq. \eqref{1} is usually referred to as the symplectic current algebra.} and can be directly used for  construction of dynamical systems on $P$. An important class of such dynamical systems was introduced in \cite{KS95,KS96} and the members of this class are called the $\E$-models. Each $\E$-model is by definition a dynamical system on the symplectic manifold $P$ defined by the Hamiltonian $H_\E$ :
\be H_\E:=\jp\int d\sigma(j(\sigma),\E j(\sigma))_\D,\label{3}\ee
where $\E:\D\to\D$ is a self-adjoint $\mathbb R$-linear involution, i.e.
\be (\E x,y)_\D=(x,\E y)_D,\quad \forall x,y\in \D; \qquad \E^2x=x,\quad \forall x\in\D.\label{4}\ee
Of course the time evolution of the current $j(\sigma)$ is obtained by calculating its Poisson bracket with the Hamiltonian which gives
\be \partial_\tau j=\partial_\sigma(\E j)+[\E j,j].\label{5}\ee 

 It turns out that every $\E$-model encapsulates the first order Hamiltonian dynamics of some nonlinear $\sigma$-model provided that it exists a Lie subalgebra ${\mathcal H}\subset\D$ the dimension of which is one half of the dimension of $\D$ and which satisfies the so called 
 isotropy condition $(x,x)_{\D}=0$, $\forall x\in{\mathcal{H}}$. 
Denoting respectively $D$ and $H$ the (simply connected) groups corresponding to the Lie algebras $\D$ and ${\mathcal{H}}$, the target space of the $\sigma$-model is the space of the right cosets
$D/H$ and its action is given by the formula \cite{KS97,KS97b}
$$ S_{\E}[f]=\frac{1}{4}\int d\tau  \oint (  f^{-1}\partial_+f,f^{-1}\partial_-f)_{\D} +  \frac{1}{4}\int d^{-1}\oint( f^{-1}df,[f^{-1}\partial_\sigma f,f^{-1}df])_{\D}- $$ \be 
  -\jp\int d\tau \oint (P_f(\E)  f^{-1}\partial_+f,f^{-1}\partial_-f)_{\D}.\label{-8b}\ee
Here the $\sigma$-model field $f(\tau,\sigma)\in {\rm Image}(\Gamma)$, where $\Gamma:D/H\to D$ is  some  fixed global section of the total fibration $D$ over the base space  $D/H$ (the ideal choice of $\Gamma$ is some global section but if there is no global section of this fibration we can choose several local sections covering the base space) and 
$P_f(\E) $ is a projection operator from $\D$ to $\D$
defined by   its kernel and its image:
\be {\rm Ker}P_f(\E)=({\rm Id}+{\rm Ad}_{f^{-1}}\E{\rm Ad}_{f})\D,\quad {\rm Image}P_f(\E)={\mathcal{H}}.\ee
The formula  \eqref{-8b} for the $\sigma$-model action from  the  $\E$-model data  $(P_\D,H_\E)$  was derived in \cite{KS97,KS97b}  and  the symplectomorphism associating to every solution of the equation of motion of the $\sigma$-model \eqref{-8b}  the solution of the first order  equation of motion \eqref{5} was given explicitely in
\cite{K15}.
 It reads
\be j =\partial_\sigma ff^{-1} -\jp f\left(P_f(\E)f^{-1}\partial_+f-P_f(-\E)f^{-1}\partial_-f\right)f^{-1}.\label{nova}\ee
If we parametrize the $\D$-valued current $j(\sigma)$ in terms of a $D$-valued variable $l(\sigma)$ as
\be j(\sigma)=\partial_\sigma l(\sigma)l(\sigma)^{-1}\ee
then the equation of motion \eqref{5} can be rewritten as
\be \partial_\tau ll^{-1}=\E\partial_\sigma ll^{-1}\ee
and we can therefore infer a useful formula
\be \partial_\pm ll^{-1}= (\E\pm {\rm Id})j.\label{usf}\ee


\section{$\E$-model underlying the Yang-Baxter $\sigma$-model with WZNW term}
We start by recalling the important concept of the Yang-Baxter operator $R:\G\to\G$, where $\G$ is the Lie algebra of a real simple  compact group $G$.  Considering  the standard Cartan-Weyl basis $H^\mu, E^\alpha$ of the complexified algebra $\G^\bc$, we introduce the basis of the real algebra $\G$ as
\be T^\mu=\ri H^\mu,\quad B^\alpha=\frac{\ri}{\sqrt{2}}(E^\alpha+E^{-\alpha}), \quad C^\alpha=\frac{1}{\sqrt{2}}(E^\alpha-E^{-\alpha}),\label{13}\ee
and the operator $R$  is then given by
\be RT^\mu=0,\quad RB^\alpha=C^\alpha,\quad RC^\alpha=-B^\alpha.\label{14}\ee
 The Yang-Baxter operator verifies the following  crucial identity
\be [Rx,Ry]=R([Rx,y]+[x,Ry])+[x,y], \quad \forall x,y\in\G. \label{15}\ee

   Let us now specify which  $\E$-model   underlies the Yang-Baxter $\sigma$-model with WZNW term  \eqref{-1} via the formula \eqref{-8b}. The relevant Lie algebra $\D$ turns out to be the complexification $\G^\bc$ of the compact simple Lie algebra $\G$. (It must be stressed that $\G^\bc$ itself has to be viewed as the {\it real} Lie algebra, e.g. for $\G=su(2)$, $\G^\bc=sl(2,\bc)$, the real Lie algebra $sl(2,\bc)$ is just the six-dimensional real Lie algebra of the Lorentz group.) There is a convenient way to describe each element $z\in\G^\bc$ in terms of two elements $x,y\in\G$ as follows 
\be z=x+\ri y.\ee
The Lie algebra commutator in $\G^\bc$ is then defined in terms of the commutator of $\G$:
  \be [z_1,z_2] :=([x_1,x_2]-[y_1,y_2])+\ri([x_1,y_2]+[y_1,x_2]), \qquad z_1,z_2\in\G^\bc, \quad x_1,y_1,x_2,y_2\in\G,\ee
  where $z_j=x_j+\ri y_j$, $j=1,2$.
  
  The second building block of an $\E$-model is the choice of the bilinear form $(.,.)_\D$. To recover the action \eqref{-1}, we choose
 for $(.,.)_\D$  a bilinear form $(.,.)_{C,\rho}$  depending on two real parameters $C$ and $\rho$, which is defined in terms of the standard  Killing-Cartan form $(.,.)$ on $\G^\bc$ as follows:
  \be (z_1,z_2)_{C,\rho}:=C.{\mathcal IM}({\rm e}^{\ri\rho}z_1z_2)\equiv C\sin{\rho}((x_1,x_2)-(y_1,y_2))+C\cos{\rho}((x_1,y_2)+(x_2,y_1)). \label{6}\ee
  Here ${\mathcal IM}$ stands for the imaginary part of a complex number.
  
The last thing to specify is the self-adjoint involution $\E:\D\to \D$ leading to the model $\eqref{-1}$ via the formula \eqref{-8b}. For that, we parametrize unambiguously any element $z\in\G^\bc$ in terms of two elements $\R,\J$ of $\G$ as follows
\be z=\frac{{\rm e}^{-\ri\rho}\R+(\cosh{p}+{\rm e}^{-\ri\rho}\sinh{p})\ri \J}{2\cosh{p}(\sinh{p}+\cos{\rho}\cosh{p})},\label{7b}\ee
where $p$ is some supplementary real parameter.
Then the operator $\E_{p,\rho}:\D\to \D$ is just defined by the flip $\R\leftrightarrow\J$
\be \E_{p,\rho}z:=\frac{{\rm e}^{-\ri\rho}\J+(\cosh{p}+{\rm e}^{-\ri\rho}\sinh{p})\ri \R}{2\cosh{p}(\sinh{p}+\cos{\rho}\cosh{p})},\label{flip}\ee
which makes its involutivity manifest. Some work is needed to verify the self-adjointness of  $\E_{p,\rho}$, i.e. the property 
\be (\E_{p,\rho}z_1,z_2)_{C,\rho}=(z_1,\E_{p,\rho}z_2)_{C,\rho}.\ee
(Note also that for $p=\rho=0$ the parametrization \eqref{7b} becomes $z=\R+\ri\J$ which explains the notation: $\R$ and $\J$ are, respectively, the "real" and the "imaginary" parts of $z$.)

\medskip

To extract from our $\E$-model $(\E_{p,\rho},\G^\bc,(.,.)_{C,\rho})$  the $\sigma$-model action $\eqref{-8b}$ we need three more things: to identify the isotropic subalgebra ${\mathcal H}\subset\G^\bc$ and the corresponding subgroup $H\subset G^\bc$, to choose the section $\Gamma$ of the $H$-fibration over $G^\bc/H$ and to find  explicit formula for the projection $P_f(\E_{p,\rho})$. Thus for the algebra ${\mathcal{H}}$ we take the subspace of $\D=\G^\bc$ 
defined as the image of  the $\br$-linear operator $\left(R-\tan{\frac{\rho}{2}}(R^2+1)-\ri\right):\G\to \G^\bc$:
\be {\mathcal{H}}_\rho:=(R-\tan{\frac{\rho}{2}}(R^2+1)-\ri)\G,\label{8e}\ee
The fact that the subspace  ${\mathcal{H}}_\rho$ is indeed a Lie subalgebra can be deduced from the identity
$$\left[\left(R-\tan{\frac{\rho}{2}}(R^2+1)-\ri\right)x,\left(R-\tan{\frac{\rho}{2}}(R^2+1)-\ri\right)y\right]=$$
\be =\left(R-\tan{\frac{\rho}{2}}(R^2+1)-\ri\right)\left(\left[\left(R-\tan{\frac{\rho}{2}}(R^2+1)\right)x,y\right]+\left[x,\left(R-\tan{\frac{\rho}{2}}(R^2+1)\right)y\right]\right)\ee which is the consequence of the identity \eqref{15}. The crucial isotropy property of  ${\mathcal{H}}_\rho$ follows from the following anti-selfadjointness property of the operator $R$ with respect to the Killing-Cartan form
\be (Rx,y)=-(x,Ry).\ee
How the group $H_\rho$ looks like? In fact, it is a  semi-direct product of some real form of the complex Cartan torus $T^\bc\subset G^\bc$ with the nilpotent group $N$ featuring the Iwasawa decomposition of $G^\bc=GAN$. To see it, it is convenient first to use the explicit form \eqref{14} of the Yang-Baxter operator $R$ to figure out that the subalgebra ${\mathcal H}_\rho\subset \D$ can be written as a semi-direct sum of two Lie subalgebras of 
$\G^\bc$ (the ideal is on the right)
\be {\mathcal H}_\rho={\rm Span_\br}( {\rm e}^{-\frac{\ri\rho}{2}}H^\mu)\subplus{\ } {\rm Span_\bc}( E^\alpha)\equiv {\rm e}^{-\frac{\ri\rho}{2}}{\rm Lie}(A) \subplus {\rm Lie}(N).\ee
Thus the subgroup $H_\rho$ of $G^\bc$ is the semidirect product $H_\rho:= A_\rho\ltimes N$, where $A_\rho$ is the Abelian  subgroup of $G^\bc$ the Lie algebra
of which is $ {\rm Lie}(A_\rho)={\rm e}^{-\frac{\ri\rho}{2}}{\rm Lie}(A)$. In the case of the group $G^\bc=SL(n,\bc)$, the elements of $A_\rho$ are
  appropriate diagonal matrices and the elements of $N$ are complex upper-triangular matrices with units on the diagonal.

If ${\rm e}^{ \frac{\ri\rho}{2}}\neq \pm\ri$ (which we are going to suppose for the rest of this paper), there exists a global section of the $H_\rho$-fibration
over $G/H_\rho$. Indeed,  this  can be deduced from the validity of the standard Iwasawa decomposition $G^\bc=GAN$ which implies in turn the validity of the modified
Iwasawa decomposition $G^\bc=GA_\rho N$. This means that any right $H_\rho$-coset in $G^\bc$ can be uniquely represented by some element of the compact group
$G$ which is, by definition, the image of that coset by the section $\Gamma$.  The section $\Gamma$ defined in this way has for the image the submanifold $G$ of $G^\bc$ therefore  the field $f(\tau,\sigma)$ in the $\sigma$-model action \eqref{-8b} can be taken as $G$-valued and denoted as $g(\tau,\sigma)$. 

In order to evaluate the   $\sigma$-model action \eqref{-8b} corresponding to the $\E$-model $(\E_{p,\rho},\G^\bc,(.,.)_{C,\rho})$, it remains to make explicit the expression $P_f(\E_{p,\rho})g^{-1}\partial_+g$. To do that we first check that  

\be g^{-1}\partial_+g= ({\rm e}^{ -\ri\rho} +\ri\cosh{p}+\ri{\rm e}^{- \ri\rho}\sinh{p})  u+(R-\tan{\frac{\rho}{2}}(R^2+1)-\ri)v, \label{Kielce} \ee
where $u,v$ are the elements of $\G$ given by the formulae
\be u=\left(\cos{\rho}+\sin{\rho}\sinh{p}+\left(R-\tan{\frac{\rho}{2}}(R^2+1)\right)(\cosh{p}+\cos{\rho}\sinh{p}-\sin{\rho})\right)^{-1}g^{-1}\partial_+g;\ee
\be v=\left(R-\tan{\frac{\rho}{2}}(R^2+1) -\frac{{\rm e}^p\tan{\frac{\rho}{2}}+1}{\tan{\frac{\rho}{2}}-{\rm e}^p}
\right)^{-1}g^{-1}\partial_+g.\ee
We note that  the $u$- and $v$-terms on the right hand side of Eq. \eqref{Kielce}  are respectively the elements of the subspaces 
${\rm Ker}P_f(\E)=({\rm Id}+{\rm Ad}_{f^{-1}}\E{\rm Ad}_{f})\D$ and ${\rm Image}P_f(\E)=\H_\rho$. This means that it holds
\be P_f(\E_{p,\rho})g^{-1}\partial_+g=\biggl(R-\tan{\frac{\rho}{2}}(R^2+1)-\ri\biggr)\left(R-\tan{\frac{\rho}{2}}(R^2+1) -\frac{{\rm e}^p\tan{\frac{\rho}{2}}+1}{\tan{\frac{\rho}{2}}-{\rm e}^p}
\right)^{-1}g^{-1}\partial_+g.\label{Presov}\ee
Inserting the expression \eqref{Presov} in the general formula \eqref{-8b}, we obtain the second order $\sigma$-model action  corresponding to the $\E$-model $(\E_{p,\rho},\G^\bc,(.,.)_{C,\rho})$: 
$$ S_{\E_{p,\rho}}[g]=\frac{C\sin{\rho}}{4}\oint\left(\int d\tau (  g^{-1}\partial_+g,g^{-1}\partial_-g)  +\int d^{-1}( g^{-1}dg,[g^{-1}\partial_\sigma g,g^{-1}dg])\right)+ $$\be+
  \frac{C}{2}\int d\tau \oint \left(\left(\cos{\rho}-\sin{\rho}\left(R-\tan{\frac{\rho}{2}}(R^2+1)\right)\right)\left(R-\tan{\frac{\rho}{2}}(R^2+1) -\frac{{\rm e}^p\tan{\frac{\rho}{2}}+1}{\tan{\frac{\rho}{2}}-{\rm e}^p}
\right)^{-1} g^{-1}\partial_+g,g^{-1}\partial_-g\right).\label{-88}\ee
The $\sigma$-model \eqref{-88} does not look like the Yang-Baxter $\sigma$-model with WZNW term as given by Eq.\eqref{-1}, however, it can be rewritten in that form by using  the identity
\be R^3=-R\ee
which permits to invert easily the operator 
$\left(R-\tan{\frac{\rho}{2}}(R^2+1) -\frac{{\rm e}^p\tan{\frac{\rho}{2}}+1}{\tan{\frac{\rho}{2}}-{\rm e}^p}
\right)$. The resulting expression reads
$$ S_{\E_{p,\rho}}[g]=\frac{C}{4\cosh{p}}\oint\left(\int d\tau (g^{-1}\partial_+g,g^{-1}\partial_-g)+ \sin{\rho}\cosh{p}    \int d^{-1}  (dgg^{-1}\stackrel{\wedge}{,}[\partial_\sigma gg^{-1},dgg^{-1}])\right)+$$
\be+\frac{C}{4\cosh{p}}\left({\rm e}^{2p}\cos^2{\frac{\rho}{2}}-\sin^2{\frac{\rho}{2}} \right)
\int d\tau\oint \left((-{\rm e}^{-p}R+R^2+1)g^{-1}\partial_+g,g^{-1}\partial_-g)\right)\label{res}\ee
and this does coincide with the original action \eqref{-1} of Delduc, Magro and Vicedo upon the identification
\be K= -\frac{C}{\cosh{p}},\quad k=\sin{\rho}\cosh{p},\quad \eta^2={\rm e}^{2p}\cos^2{\frac{\rho}{2}}-\sin^2{\frac{\rho}{2}}, \quad
A=-{\rm e}^{p}\cos^2{\frac{\rho}{2}}+{\rm e}^{-p}\sin^2{\frac{\rho}{2}}. \label{Uh}\ee
It can be also checked that the constraint \eqref{-2} is satisfied by the values of the parameters $K,k,\eta,A$ given by Eqs. \eqref{Uh},
which means that the crucial property of integrability of the Yang-Baxter $\sigma$-model with WZNW term comes about automatically from our $\E$-model formalism.  

\section{Integrability}

The first order field equations \eqref{5} of  the $\E$-model defined by the data $(\E_{p,\rho},\G^\bc,(.,.)_{C,\rho})$ acquire a particularly simple form if we parametrize the current $j(\tau,\sigma)$ as in Eq. \eqref{7b}
  \be j=\frac{{\rm e}^{-\ri\rho}\R+(\cosh{p}+{\rm e}^{-\ri\rho}\sinh{p})\ri \J}{2\cosh{p}(\sinh{p}+\cos{\rho}\cosh{p})}.\label{7e}\ee
  We then have
  \be \E_{p,\rho}j=\frac{{\rm e}^{-\ri\rho}\J+(\cosh{p}+{\rm e}^{-\ri\rho}\sinh{p})\ri \R}{2\cosh{p}(\sinh{p}+\cos{\rho}\cosh{p})}\label{7f}\ee
  therefore Eq.  \eqref{5}  for the data $(\E_{p,\rho},\G^\bc,(.,.)_{C,\rho})$ can be equivalently rewritten in terms of two $\G$-valued equations :
  \begin{subequations}\label{y}
\begin{align}
\label{ 7c}
 \partial_\tau \R&=\partial_\sigma\J+[\J,\R]; 
 \\
\label{ 7d}
 \partial_\tau \J&=\partial_\sigma\R. 
\end{align}
\end{subequations}
The form \eqref{y} of the first order field equations is of course very familiar since it is identical with the Zakharov-Mikhailov field equations of the principal chiral model \cite{ZM}  allowing the so called Lax pair with spectral parameter. This means that given a solution $\R(\tau,\sigma),\J(\tau,\sigma)$ 
of the equations of motion \eqref{y} and a complex number $\lambda\neq\pm 1$ the following $\G^\bc$-valued  zero curvature identity takes place
\be \partial_+A_-(\lambda)-
\partial_-A_+(\lambda)+\left[A_-(\lambda), A_+(\lambda)\right]=0, \label{9}\ee
where
\be A_\pm(\lambda):=\frac{\J\pm\R}{1\pm\lambda}.\label{9b}\ee
We thus conclude that our $\E$-model $(\E_{p,\rho},\G^\bc,(.,.)_{C,\rho})$ is integrable, in the sense of admitting an infinite number of conserved observables extracted from the Lax pair $A_\pm(\lambda)$ in the standard way \cite{BBT}.
We   have  thus given an alternative proof of integrability  of the model \eqref{-1} than that presented in \cite{DMV15}.

We remark that  the Yang-Baxter $\sigma$-model with WZNW term     was shown to be integrable even in the stronger sense in \cite{DMV15} where it was demonstrated via the so called $r/s$-formalism  that the conserved observables extracted from the Lax pair Poisson-commute. The matrix Poisson brackets of the components of the Lax connection  which intervene in this computation do depend on the parameters  $C,p$ and $\rho$, which is not surprising since the Poisson brackets of the currents $\R$ and $\J$   depend on them. Indeed we have

  \begin{subequations}\label{les}
\begin{align}
\{\J^a(\sigma_1),\J^b(\sigma_2)\}&=\phantom{-}\frac{2\cosh{p}}{C}
f^{ab}_{\phantom{ab}c}\J^c(\sigma_1)\delta(\sigma_1-\sigma_2) -\frac{4(\cosh{p})^2\sin{\rho}}{C}\delta^{ab}\delta'(\sigma_1-\sigma_2);
 \\
\{\J^a(\sigma_1),\R^b(\sigma_2)\}&=\phantom{-}\frac{2\cosh{p}}{C}f^{ab}_{\phantom{ab}c}\R^c(\sigma_1)\delta(\sigma_1-\sigma_2) -\frac{4(\cosh{p})^2(\cosh{p}+\cos{\rho}\sinh{p})}{C}\delta^{ab}\delta'(\sigma_1-\sigma_2);
 \\
\{\R^a(\sigma_1),\R^b(\sigma_2)\} &=-\frac{2\cosh{p}\left((\sinh{p})^2+(\cosh{p})^2+2\cos{\rho}\sinh{p}\cosh{p}\right)}{C}\J^c(\sigma_1) f^{ab}_{\phantom{ab}c}\delta(\sigma_1-\sigma_2)+\\
&~~~~~~~~~~~~~~+\frac{4\sin{\rho}(\cosh{p})^2}{C}
 \R^c(\sigma_1)f^{ab}_{\phantom{ab}c}\delta(\sigma_1-\sigma_2)   
 -\frac{4(\cosh{p})^2\sin{\rho}}{C}\delta^{ab}\delta'(\sigma_1-\sigma_2).
\end{align}
\end{subequations}
The brackets \eqref{les} were obtained from the symplectic current algebra \eqref{1} in the following way:
  we chose a basis $T^A:=(\tau^a,\ri\tau^a)$ of $\G^\bc$  where  $\tau^a$ is an orthonormal basis of $\G$ satisfying  
  \be (\tau^a,\tau^b)=-\delta^{ab}, \quad [\tau^a,\tau^b]=f^{ab}_{\phantom{ab}c}\tau^c,\ee
  we  set
\be \R^a:=(\R,\tau^a),\quad \J^a:=(\J, \tau^b)\ee
and, using the parametrization \eqref{7e}  of the $\D$-current $j$, we evaluated
\be (j,\tau^a)_{C,\rho}=\frac{C\J^a}{2\cosh{p}}, \quad (j,(\sinh{p}+{\rm e}^{-\ri\rho}\cosh{p})\ri\tau^a)_{C,\rho}=\frac{C\R^a}{2\cosh{p}}. \label{Lem}\ee
Finally we plugged the expressions \eqref{Lem} into  the symplectic current algebra \eqref{1}. 

For completeness, we give also the formula for the  Hamiltonian  \eqref{3} of our $\E_{p,\rho}$-model in the $\R,\J$-parametrization \eqref{7e}    
\be H_{\E_{p,\rho}}:=\frac{C}{2}\int d\sigma\frac{(\cosh{p}+\cos{\rho}\sinh{p})((\R,\R)+(\J,\J))-2\sin{\rho}(\R,\J)}{4(\cosh{p})^2(\sinh{p}+\cos{\rho}\cosh{p})^2}.\label{7g}\ee
Combining the formulae \eqref{les} and \eqref{7g} we find for the Poisson brackets $\{\R,H_{\E_{p,\rho}}\}$ and $\{\J, H_{\E_{p,\rho}}\}$
\be\{\R,H_{\E_{p,\rho}}\}= \partial_\sigma \J +[\J,\R]; \quad \{\J, H_{\E_{p,\rho}}\}= \partial_\sigma  \R\ee
in full accord with the equations of motion \eqref{y}. 

\section{Another form of the first order formalism}

The reader might wish to understand better how the second order action \eqref{res} can be derived from the symplectic current algebra \eqref{les} and the Hamiltonian \eqref{7g}.  In a general case, they  may wish to consult the original papers \cite{KS95,KS97,KS97b}, but the case of the 
Yang-Baxter $\sigma$-model with the WZNW term given by \eqref{res} can be understood quite simply by suitably parametrizing  the symplectic current algebra in terms of the coordinates of another symplectic manifold which features in the first order description of the dynamics of nonlinear $\sigma$-models with the WZNW terms. Consider thus the spaces $L\G$ and $LG$ consisting of smooth maps from the circle $S^1$ respectively into the Lie algebra $\G$ and into  the Lie group $G$. The manifold $LG\times L\G$ is referred to as the (co)tangent bundle of the loop group $LG$ and it is equipped with a   symplectic structure\footnote{For $k=0$ the expressions \eqref{Ws} describe the canonical symplectic structure on the (co)tangent bundle.}
$\Omega$  
given by two equivalent expressions:
 \begin{subequations}\label{Ws}
\begin{align}
\label{Wsa}
 \Omega_{}&=\phantom{-}\frac{k}{2}\oint(dgg^{-1}\stackrel{\wedge}{,}
\partial_\sigma(dgg^{-1}))+d\oint(dgg^{-1},J_L), \qquad (g,J_L)\in LG\times L\G;
 \\
\label{Wsb}
\Omega_{}&=-\frac{k}{2}\oint( g^{-1}dg\stackrel{\wedge}{,}
\partial_\sigma( g^{-1}dg))-d\oint( g^{-1}dg,J_R), \quad J_R=-g^{-1}J_Lg+kg^{-1}\partial_\sigma g.
\end{align}
\end{subequations}
Here the symbol $\oint$ stands for the circular integral around the loop parameter $\sigma$     and $(.,.)$ is the Killing-Cartan form on $\G$.

It is useful to recall the Poisson brackets of the variables $g,J_L,J_R$ induced by the symplectic form \eqref{Ws} that were computed e.g. in \cite{K04,V15}:
 \begin{subequations}\label{WPB}
 \begin{align}
\label{WPBa}
 \{g(\sigma_1)\stackrel{\otimes}{,}g(\sigma_2)\}&=0;
\\
\label{WPBb}
\{g(\sigma_1),(J_L(\sigma_2),\tau^a)\}&=\phantom{-}\tau^a g(\sigma_1)\delta(\sigma_1-\sigma_2);
\\
\label{WPBc}
\{g(\sigma_1),(J_R(\sigma_2),\tau^a)\}&=-g(\sigma_1)\tau^a\delta(\sigma_1-\sigma_2); 
\\
\label{WPBd}
 \{(J_L(\sigma_1),\tau^a),(J_L(\sigma_2),\tau^b)\}&= (J_L(\sigma_1),[\tau^a,\tau^b])\delta(\sigma_1-\sigma_2)+k(\tau^a,\tau^b)\delta'(\sigma_1-\sigma_2);
 \\
 \label{WPBe}
 \{(J_R(\sigma_1),\tau^a),(J_R(\sigma_2),\tau^b)\}&= (J_R(\sigma_1),[\tau^a,\tau^b])\delta(\sigma_1-\sigma_2)-k(\tau^a,\tau^b)\delta'(\sigma_1-\sigma_2);
 \\
 \label{WPBf}
 \{(J_L(\sigma_1),\tau^a),(J_R(\sigma_2),\tau^b)\}&=0.
\end{align}
\end{subequations}

Consider now an injective map $\Upsilon:LG\times L\G\to P$ from the cotangent bundle of the loop group $LG$ into the symplectic current algebra $P$ (based on the choice $\D=\G^\bc$) given explicitely by
\be j(\sigma)\equiv\Upsilon(g,J_L):=\partial_\sigma gg^{-1}+\frac{e^{-\ri \rho}}{C}(R_{g^{-1}}+\tan{\frac{\rho}{2}}(R_{g^{-1}}^2+1)-\ri)(-J_L+C\sin{\rho} \ \partial_\sigma gg^{-1}),\label{14b}\ee
where
\be R_{g^{-1}}:= {\rm Ad}g R  {\rm Ad}g^{-1}.\ee
The crux of the story is the fact that, for the choice $k=C\sin{\rho}$ and for the choice of the bilinear form  $(.,.)_\D=(.,.)_{C,\rho}$   given by \eqref{6},
the map $\Upsilon$ preserves
the Poisson brackets \eqref{WPB} on  $LG\times L\G$ and \eqref{1} on $P$ (as it can be verified by somewhat tedious calculation employing crucially the Yang-Baxter identity \eqref{15}). The procedure of the extraction of a non-linear $\sigma$-model \eqref{res} from the $\E_{p,\rho}$-model based on the involution \eqref{flip} then starts by pull-backing the symplecting form $\omega$  and  the Hamiltonian $H_{\E_{p,\rho}}$ 
from $P$ to $LG\times L\G$ where they become, respectively,  $\Upsilon^*\omega=\Omega$ and $\Upsilon^*H_{\E_{p,\rho}}=H_{p,\rho}$ :
\be H_{p,\rho}(g,J_L)=\jp\oint(\Upsilon(g,J_L),\E_{p,\rho}\Upsilon(g,J_L))_{C,\rho}.\ee
It is well-known how to obtain the first order action of a general dynamical system $(P,\omega,H)$ if the symplectic form $\omega$ were a coboundary, that is, if  it existed a $1$-form $\theta$ such that $\omega=d\theta$. In this case the action $S(\gamma)$ of a trajectory $\gamma(\tau)$ in the
phase space $P$ would be given by
\be S(\gamma)=\int_{\tau_i}^{\tau_f} d\tau ( \gamma^*\theta - (\gamma^*H) d\tau ).\ee
If $\omega$ is not coboundary, we must write instead
\be S(\gamma)=\int_{\tau_i}^{\tau_f} d\tau ( \gamma^*d^{-1}\omega - (\gamma^*H) d\tau).\label{fob}\ee
Strictly speaking, the object $d^{-1}\omega$ is not well-defined but  its variation is, and that is all what is needed to extract field equations
out from \eqref{fob}. We do not enter the discussion how precisely the term $d^{-1}\omega$ should be handled in general (this is a well-known story for everyone who is familiar   with the dynamics of the standard WZNW model)    since what  matters for us is that the 
symplectic form  $\Omega$ given by \eqref{Wsa} is the sum of two $2$-forms from which the one containing explicitly the variable $J_L$ is the coboundary. This
permits us to write the first order action of a trajectory $\gamma(\tau):=(g(\tau),J_L(\tau))$ of the dynamical system $(LG\times L\G, \Omega, H_{p,\rho})$ as follows
$$S_{p,\rho}=\int d\tau  \oint \left ((\partial_\tau gg^{-1},J_L) -\jp(\Upsilon(g,J_L),\E_{p,\rho}\Upsilon(g,J_L))_{C,\rho}\right) + \frac{C\sin{\rho}}{2}\int \gamma^*d^{-1}\oint(dgg^{-1}\stackrel{\wedge}{,}
\partial_\sigma(dgg^{-1}))=$$
$$ = \int d\tau \oint \left((\partial_\tau gg^{-1},J_L) -\frac{e^{-p}}{C} (J_L,J_L)-\frac{\sinh{p}+\cos{\rho}\cosh{p}}{C(1+\cos{\rho})} (R_{g^{-1}}J_L,R_{g^{-1}}J_L)\right)+$$
$$+\int d\tau \oint  \left(\partial_\sigma gg^{-1},\left(\sin{\rho}\cosh{p}+(\sinh{p}+\cos{\rho}\cosh{p})\left(R_{g^{-1}} -\tan{\frac{\rho}{2}}
(R^2_{g^{-1}}+1)\right)\right)J_L\right) $$
\be -\frac{C(\cosh{p}+\cos{\rho}\sinh{p})}{2} \int d\tau \oint   (\partial_\sigma gg^{-1},   \partial_\sigma gg^{-1}) + \frac{C\sin{\rho}}{2}\int \gamma^*d^{-1}\oint(dgg^{-1}\stackrel{\wedge}{,}
\partial_\sigma(dgg^{-1})).
 \label{foc2}\ee
 We note that the $\G$-valued current $J_L$ plays the role of auxiliary field in \eqref{foc2} and, moreover, it appears there quadratically. It can be therefore easily eliminated giving a second order action
 $$S_{p,\rho}=\frac{C    e^p(1+\cos{\rho})}{4}\int d\tau \oint(g^{-1}\partial_+g,   g^{-1}\partial_- g )+$$$$ 
 + \frac{C}{4}\left(\frac{1}{\cosh{p}}-e^p(1+\cos{\rho})\right) \int d\tau \oint  (Rg^{-1}\partial_+g,   Rg^{-1}\partial_- g ) +$$
 \be +\frac{C(\sinh{p}+\cos{\rho}\cosh{p})}{2\cosh{p}}\int d\tau \oint  (Rg^{-1}\partial_\tau g,   g^{-1}\partial_\sigma g )
 +\frac{C\sin{\rho}}{4}\int d^{-1}\oint (dgg^{-1}\stackrel{\wedge}{,}[\partial_\sigma gg^{-1},dgg^{-1}]).\label{foc3}\ee
Of course, the action \eqref{foc3} can be easily rewritten in the equivalent form \eqref{res} of the Yang-Baxter $\sigma$-model with WZNW terms, as it should.



\section{Outlook}

The ordinary Yang-Baxter $\sigma$-model \cite{K02,K09}  obtained from \eqref{res} by setting $\rho=0$ can be alternatively deformed to  the two-parametric bi-Yang-Baxter $\sigma$-model \cite{K14} which is also integrable.
There is an obvious open problem of "switching on"  a non-vanishing $\rho$ in the bi-Yang-Baxter case and to obtain in this way a three-parametric integrable deformation of the principal chiral model.  This problem does not seem to be an easy one, however.




\end{document}